# Femtosecond and Ultraviolet Laser Irradiation of Graphite-Like Hexagonal Boron Nitride

Andrei V. Kanaev[#] and Jean-Pierre Petitet

*Laboratoire d'Ingénierie des Matériaux et des Hautes Pressions – LIMHP, CNRS, Institut Galilée, Université Paris 13, 93430 Villetaneuse, France*

Luc Museur

*Laboratoire de Physique des Lasers – LPL, CNRS, Institut Galilée, Univerité Paris 13, 93430 Villetaneuse, France*

Vladimir Marine

*Groupe de Physique des Etats Condensée – GPEC, CNRS, Université Marseille, 13288 Marseille, France*

Vladimir L. Solozhenko[1]

*Laboratoire des Propriétés Mécaniques et Thermodynamiques des Matériaux – LPMTM, CNRS, Institut Galilée, Université Paris 13, 93430 Villetaneuse, France*

Vassilis Zafiropulos[2]

*Laser & Applications Division, Institute of Electronic Structure & Laser (I.E.S.L.), Foundation for Research and Technology - Hellas (F.O.R.T.H.), Heraklion 71110, Crete, Greece*

[#] corresponding author. E-mail: kanaev@limhp.univ-paris13.fr
[1] on leave from the Institute for Superhard Materials Ac. Sci., Kiev, Ukraine
[2] present address: Laboratory of Applied Physics, Department of Human Nutrition & Dietetics, Technological Educational Institute of Crete, Ioannou Kondylaki 46, 723 00 Sitia, Crete, Greece.

**Abstract**

Effect of the femtosecond and nanosecond UV laser irradiation (below the ablation threshold) of graphite-like hexagonal boron nitride (hBN) has been studied. Experiments were carried out with the compacted powder under high vacuum at room temperature using excimer KrF laser (248 nm). In the nanosecond operation mode, the laser-induced fluorescence spectra are found strongly modified depending on the integrated doze, which is attributed to a progressive enrichment of the surface layer by elemental boron. A slow sample recovery after the laser irradiation has been observed. On the other hand, in the femtosecond mode the fluorescence spectra depend on the laser fluence, and the changes are reversible: low energy fluorescence spectra are restored immediately when the laser energy decreases. This effect can be explained by a material bleaching, which favors a bulk centers emission. The ablation threshold has been determined as 78 mJ/cm² in the femtosecond laser operational mode.

## I. INTRODUCTION

Many experimental and theoretical activities have been devoted in recent years to studies of physical and chemical properties of boron nitride (BN).[1,2] Besides amorphous state (aBN), four crystalline polymorphs of BN are known: hexagonal (hBN), rhombohedral (rBN), cubic (cBN) and wurzitic (wBN) ones. They correspond to hexagonal and rhombohedral graphite, diamond, and hexagonal diamond, respectively. The hBN is the most commonly used form of boron nitride, provides a basis for many advanced technologies. A considerable interest to another family member – cubic BN – is due to its extreme hardness putting this material on the second place after diamond for industrial applications. Useful properties include also thermal and chemical stability, high melting temperature, wide band gap, high thermal conductivity, and low dielectric constant. All together, this makes cBN important in many commercial applications: superabrasives, microelectronic devices, protective coatings, etc.

As it has been shown by Solozhenko,[3] cBN is thermodynamically stable boron nitride polymorph at ambient conditions. However, a direct solid-state transformation of hBN into cBN requires overcoming a considerable activation barrier. Because of this the hBN-to-cBN transformation occurs only at temperatures above 2000 K and pressures exceeding 11 Gpa.[4] By use of traditional solvents, the p,T-parameters of cBN formation can be reduced down to 4 GPa and ~1300 K.[5] Somewhat lower p,T-conditions has been reported for cBN formation from amorphous or nanocrystalline boron nitride,[1] however, the mechanism of cBN nucleation is not clear.

A laser processing of hBN may be considered as an alternative approach to achieve the h→c phase transformation. The effect of the laser radiation on boron nitride has been previously studied using IR Nd:YAG (1.064 μm)[6] and UV $N_2$-laser (337.1 nm)[7] of microsecond pulse duration, and a possibility of hexagonal-to-cubic phase transition has been reported. Thermal effects would play an important role in this transformation. Although the mechanism has not been understood, because of a relatively low power density (<400 kW/cm$^2$), participation of the "soft" photochemical mechanism has to be considered.

In the present communication we report on the effect of the femtosecond and nanosecond UV laser irradiation (KrF, excimer 248 nm) of graphite-like hexagonal boron nitride (hBN) at energy densities below the ablation threshold. A comparative study has been carried out using in situ fluorescence spectroscopic technique.

## II. EXPERIMENTAL SETUP

Experiments were carried out at the I.E.S.L.-F.O.R.T.H. using nanosecond ($t_{L1}$=15 ns) and femtosecond ($t_{L2}$=450 fs) excimer KrF lasers (248 nm, 10 Hz, $F_L$=1-100 mJ/cm²/pulse). The ultrashort pulses were generated using a XeCl excimer pumped dye laser system producing 450 fs pulses from a distributed feedback dye laser (DFDL) at 496 nm, which after frequency doubling were amplified in a KrF excimer laser cavity (Lambda-Physik). The maximum pulse energy was ~ 10 mJ.

The laser beam was focused on the sample maintained on a stainless holder inside a vacuum chamber (~ $10^{-7}$ mbar) at room temperature. The almost homogeneously illuminated rectangular-like spot has been produced on samples. The energy density on the sample has been carefully measured using an energy meter and taking into account the beam geometry. A typical energy density values for the fluorescence experiments were around several mJ/cm². The emission induced by the excitation beam was collected by a quartz lens and focused on a quartz optical fiber. An X-Y micrometer ensures accurate positioning of the optical fiber relatively to the fluorescent image. The light from the fiber was spectrally analyzed in a 0.20 m grating spectrograph equipped with a 300 grooves/mm grating, which provides an effective spectral window of 280 nm at the exit port of the spectrograph. The emission spectrum was recorded on an optical multichannel analyzer based on an intensified photodiode array detector (OMA III system, EG&G PARC Model 1406). A computer interface system was finally used for the storage and analysis of the recorded spectra. Moreover, a delay between the laser pulse and the signal recording was introduced with a step of 5 ns.

The samples were compacted from hBN powder (Alfa, 99,5%) under the pressure of 0.6 GPa in square pallets (8x8x1 mm). To avoid organic impurities and traces of water, the pallets were heated at 800 K during 12 hours. The grit size of the hBN powder used for the experiment has been estimated by granulometry and transmission electron microscopy (JEM 100C JEOL). It ranged from 0.3 to 10 μm with an average particle size of 3.1 μm corresponding to the maximum in the mass distribution curve. The specific surface value was 2.6 m$^2$/g estimated from the absorption isotherm of nitrogen at 77 K by the BET-method. The impurity contents of the hBN sample under study according to the data of mass-spectrometry, spectral analysis and X-ray microanalysis are given in Table I. Using conventional X-ray powder diffraction, the lattice parameters of hBN were determined to be $a = 2.504(2)$ Å and $c = 6.660(8)$ Å. The degree of three-dimensional ordering of the sample studied has been calculated by the method described in Ref. 8 to be $P_3 = 0.98 \pm 0.02$.

## III. RESULTS AND DISCUSSION

The hBN samples are highly fluorescent under the optical excitation above the bandgap energy of $E_g(hBN) = 4.02 \pm 0.01$ eV ($H \rightarrow M$).[9] The fluorescence spectra of dry hBN powder show different bands: (i) asymmetric-shape continuum with a maximum at ~370 nm and FWHM=100 nm (UV1), (ii) structured band with maxima at 320, 336 and 354 nm (UV2), and (iii) broad visible continuum with maximum at ~440 nm and FWHM=150 nm (V1). The UV1 band dominates at excitation energy above 4.77 eV ($\lambda \leq 260$ nm), whereas the UV2 band appears at excitation between between 4.77 eV and the bandgap energy $E_g$ (260 nm $\leq \lambda \leq$ and 308 nm). The UV1 band has been tentatively assigned to the surface state emission, and the UV2 band was assigned to shallow traps or to m-STE centers.[9] The V1 emission band, which is not specific of excitation wavelength, might have its origin in photochemistry catalyzed on hBN particles. Below we will make use of these characteristic emissions bands for a discussion of the sample modifications.

<u>Nanosecond laser irradiation</u>

In the nanosecond experiments the laser power density has been maintained below the ablation threshold. The laser excitation corresponds to photon energies where the fluorescence spectra consist principally of UV1 with a small admission of UV2 band. Indeed, we have observed a characteristic emission, which is shown in Fig. 1 (the uppermost frame, N=1). Surprisingly, the emission spectra change the shape under the laser irradiation: the band UV2 appears when the number of laser pulses N increases (see the lower frames in Fig. 1). The SEM image (Fig. 2) shows that the surface roughening of the sample decreases during the laser treatment. In the same time visible changes in the sample color appear: it becomes slightly brown. The elemental analysis carried out by the ESCA method has shown that the laser impact region was enriched by boron. Moreover, the boron enrichment of the surface layer under the laser irradiation can be confirmed by the analysis of the plasma plume in the ablation experiments, which allows us a direct elemental probing of the surface material. In the UV-visible spectral range we have observed 3 principal peaks at 345 nm, 582.5 nm and 592 nm that can be assigned to the ionized and neutral boron and nitrogen atoms. Indeed, the $B^+$ line at 345.1287 nm, $N$ line at 582.954 nm and a group of $N^+$ lines at ~592 nm (591.033 nm, 591.407 nm, 592.722 nm, and 592.781 nm) are known.[10] In Fig. 3 we show two spectra measured for the initial and irradiated hBN samples. For simplicity, the spectra in this Figure are normalized on the $N$ line intensity. We can see that the $B^+$ line intensifies after the laser irradiation and the $N^+$ lines decrease in intensity. This is a strong indication of the sample enrichment by boron. The process starts immediately after the beginning of the sample irradiation, however it saturates with time. The sample color becomes gray but never dark, which means that deeper layers of hBN below the sample surface do not contribute to the boron production. Neither conventional Raman spectroscopy nor X-ray diffraction have indicated any structural changes in the sample. This may be related to the extremely small thickness of the transformed layer.

### Surface chemistry

The boron enrichment of the surface layer may be due to a thermal effect or a photochemical action of the UV laser photons. The first mechanism has been already considered by Doll et al [11] using ablative interaction of hBN samples with a pulsed KrF laser (248 nm). As a result of it, the surface morphology has been changed considerably, and boron reach spheroids were found to cover the irradiated area. This effect is not significant in our experiments due to a low laser power density below the ablation threshold. Indeed, as Fig. 1 shows, at low laser fluences the net effect of the irradiation is rather the surface cleaning.

Between different pathways of the photochemical hBN decomposition, only one is available with the KrF laser photons ($h\nu$ =5.0 eV) at low fluence, namely

$$BN(s) \rightarrow B(s) + 1/2 N_2(g) - 2.57 eV \,, \qquad (1)$$

where (s) and (g) represent solid and gas phases, respectively. The process (1) is responsible for the change of the sample composition under irradiation of samples with intense visible-IR laser pulses.[12] Its rate can be small close to the activation energy of 2.57 eV but is considerably increased using photonic sources above the fundamental absorption edge energy of hBN. Another reaction channels, which conserve the sample composition, are characterized by the activation energy exceeding that of the laser photons. The less energy consuming one is:

$$BN(s) \rightarrow BN(g) - 7.47 eV \,, \qquad (2)$$

The process kinetics has been studied using the fluorescence spectra. As it is seen from the spectra, the UV2 band is intensified during the laser treatment with respect to the UV1 band. This makes sense if one takes into account the band assignment. The absorbed energy can migrate in the form of excitons to the hBN surface and than be trapped. This process results in the UV1 emission. The laser irradiation may passivate the surface decreasing the number of traps. When laser beam is off, a slow recovery of the fluorescence spectra is observed, which is illustrated by Fig. 1. This is indicative of the recovery of the surface traps. Under vacuum of $10^{-7}$ mbar, oxygen molecules are sufficiently abundant species that can react with the as-formed boron on the sample surface. We therefore assume, that it is the oxide layer that is progressively formed as a dark reaction channel

$$xB(s) + \frac{y}{2}O_2(g) \rightarrow B_xO_y(s) \qquad (3)$$

The boron oxidation may continue until a layer of stable $B_2O_3$ is formed. The boron oxide surface layer may be responsible for the UV1 fluorescence band.

The sample modification can therefore be observed by following the intensity ratio $R$ of the UV2-to-UV1 band intensities as a function of the laser dose, $D_L = \sum_i F_{L_i}$ (fluence of the i-th laser of duration $\tau$ is $F_{L_i} = \int_\tau I_{L_i}(t)dt$, where $I_{L_i}$ stands for the laser intensity). This is shown in Fig. 4. We note that during the experiment the laser fluence $F_L$ was varied by a factor of 10 and the irradiation of the sample between series with different $F_L$ was several times delayed (the longest delay of $\Delta t = 1.5$ h is seen at $D_L \approx 55$ J.cm²).

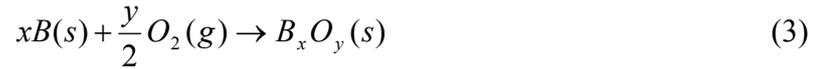

*Modeling*

To prove the photochemical nature of the sample modification, we model these results using a basic phenomenological model addressing above trends. The kinetic equation for the number of the active surface states *n* producing UV1 emission can be written as

$$\frac{dn}{dt} = -\gamma I_L n + k(n_0 - n),$$ (4)

where $n_0$ is the total number of active sites and $\gamma$ and $k$ are the constants of the dissociation process (1) and the reaction of oxidation (3), respectively. We express the normalized intensity of the UV2 band as $I = a(1 - n/n_0)$, where $a$ is a constant. The equation (4) can be solved by iterations relatively $I_i$ as a function of the time interval i: (a) after the i-th period of the laser action, and (b) after the i+1-th period between two laser pulses. This is justified by a fact that laser pulse (20 ns) is much shorter compared to the delay time between pulses (100 ms). Following this approach we have obtained two solutions of Eq. (4) for cases (a) and (b), respectively:

$$I_i = aI_{i-1} \cdot (1 - (1 - I_{i-1}/a)\exp(-\gamma F_{L_i}/h\nu))$$ (5a)

$$I_{i+1} = I_i \cdot \exp(-k\Delta t_i)$$ (5b)

where $F_L$ is the laser fluence. The theoretical curve of $I(D_L)$ calculated using two fit parameters of $k = 1.5 \cdot 10^{-4}$ s$^{-1}$ and $\gamma = 0.030$ cm²/J is shown in Fig. 4 by the solid line. Despite of a rather complex representation of $F_L$ our theoretical model is in a good agrement with the experiment. As we see, the produced effect is proportional to the number of the laser photons and it is independent on the laser light intensity. This supports our conclusion about the photochemical (non thermal) nature of the process.

Using the obtained constants we can make estimations of the reaction cross-sections. Firstly, we considered $k = 1.5 \cdot 10^{-4}$ s$^{-1}$, which according to our model belongs to the "dark-channel" reaction (3). Under the experimental condition the oxygen concentration in the vacuum chamber ($p \approx 10^{-7}$ mbar) is $n = 5 \cdot 10^8$ cm$^{-3}$. The flow rate on the target surface is $\phi = 1/4 nV \approx 5 \cdot 10^{13}$ cm$^{-2}$s$^{-1}$. The cross-section of the surface reaction (3) is, therefore, $\sigma_3 = k/\phi \approx 3.0 \cdot 10^{-18}$ cm². The effective dissociation cross-section of the photochemical surface reaction (1) is $\sigma_1 = \gamma \cdot h\nu = 2.4 \cdot 10^{-20}$ cm², which is considerably smaller than the geometric molecular cross-section. However, in this case one has to take into account the process efficiency, which can be low because of a low probability of the local phonon mode overheating resulting to dissociation.

The observed effect is attributed to the surface dissociation of hBN following by the oxidation of elemental boron in the surface layer. Boron oxide is transparent material in the visible. This explains the residual coloration of the sample surface.

## Femtosecond laser irradiation

A similar effect of the laser irradiation on hBN fluorescence spectra has been observed under femtosecond laser irradiation below ablation threshold, which measured value of ~78 mJ/cm² (in a single pulse) is typical for inorganic insulators. Surface processing by femtosecond laser pulses reveals that much lower laser energy requires observing the effect using femtosecond pulses compared to the nanosecond ones. Moreover, in this irradiation mode the effect is a function of the laser fluence in contrast to nanosecond pulses, where we observe the doze dependence. The observed spectra are presented in Fig. 5. The analysis of decay fluorescence shows the lifetimes $\tau_{UV} = 5.2 \pm 0.4$ ns and $\tau_{VIS} = 73 \pm 9$ ns for the ultraviolet (UV1+UV2) and visible (V1) bands, respectively. The lifetimes of the UV1 and UV2 bands have not been measured separately, however their difference is not strong. The relative intensity of the fluorescence UV2 band as a function of laser fluence is shown in Fig. 6. Low fluence spectra measured without delay at the end of each irradiation pulse are also shown in this Figure. We can see that the spectral changes are completely reversible: the low energy fluorescence spectra are immediately recovered when the laser energy is decreased. As it follows from our measurements, in the range of small fluences ($0.7 \le F_L \le 9.0$ mJ/cm²) the sample is not modified, although the fluorescence spectra do. Their intensity increases almost linearly with laser energy. Atomic boron may be produced at the sample surface at laser fluence value above 9.0 mJ/cm², which was observed by a slight attenuation of the integral fluorescence yield. In the same time, the spectral line shape corresponds to the recovered sample (in terms of the nanosecond laser processing). The fact indicates that in this case the net effect is related neither to the surface bond breaking nor to the surface cleaning from the oxidizing species. The sample recovery takes place during the femtosecond laser pulse. Indeed, the highest used doze in the current experimental series was 4.3 J/cm², which is much less than in the ns-laser experiments (200 J/cm²). Therefore, the above-discussed nanosecond photochemistry (1) cannot be responsible for the intensity changes.

We can now estimate an upper limit of the local surface temperature of hBN sample under the laser treatment. A small value of the dissociation cross-section $\sigma_2 / \sigma_{at} \approx 2.4 \cdot 10^{-4}$ suggests a low efficiency of the dissociation process (1). Neglecting the heat transfer we obtain the expression for the temperature change in the irradiated spot after the electronic relaxation: $\Delta T_{surf} \leq F_L \cdot \alpha / \rho C_p$, where $\alpha \approx 2 \cdot 10^5$ cm$^{-1}$ is a typical value of the interband absorption coefficient, $\rho$=2.34 g/cm$^3$ is hBN density, and $C_p \approx$0.8 J/(g K) is the heat capacity of hBN at room temperature.[13] At irradiation of laser fluence values $F_L \leq 20$ mJ/cm², the local surface temperature is below the hBN melting point, $T_{melt} \approx$3000 K.[14] At higher fluence, the local heat results in material melting and evaporation, and the ablation appears at $F_L^* = 78$ mJ/cm². In the femtosecond regime the involved process of electron absorption ($t_{L2}$=450 fs), heat transfer to the matrix ($\tau_{el}$ ~10 ps), melting, and heat conductivity are separated in time (they are cited in order from faster to longer ones). In the nanosecond laser regime both processes, absorption and heat conductivity, occur on the same timescale. Therefore, the local surface temperature is lower. From these estimations we can conclude that below the saturation of the fluorescence signal in Fig. 6 we record spectra of solid hBN samples, and thus, the spectral effect cannot be related to the melting.

Because of the long nanosecond lifetimes ($\tau_{UV} = 5.2$ ns) of the fluorescent bands (both UV1 and UV2), the difference in their behavior under *ns* and *fs* laser excitation cannot be explained by a bottleneck of the population scheme. Indeed, the UV2 band is strongly intensified at $F_L \geq 5$ mJ/cm² in the femtosecond laser regime (500 fs), whereas no change in the sample spectra has been observed at 20 times higher fluence ($F_L = 100$ mJ/cm²) in the nanosecond regime (15 ns). Moreover, the spectra are similar in both laser regimes indicating the same band origin. We keep in mind the band identification inferred from the *ns* laser experiments: UV2 – bulk states (STE), and UV1 – surface states. We believe that the intensification of the UV2 band in the fs-laser mode is due to a longer radiation penetration depth into the solid hBN. Indeed, the absorbed radiation causes excitation in the solid. This excitation in form of polarons or excitons can than be trapped either in the sample volume or at its surface. The energy transport to the surface occurs by the exciton diffusion characterized by the diffusion length $l_{exc}$. The higher is the ratio $\delta / l_{exc}$ ($\delta$ is the light penetration depth) the stronger is the probability of the bulk exciton trapping. Therefore, the increase in $\delta$ should result in an increase of the UV2 emission intensity relative to the UV1 one. Below discuss this point.

### Bleaching mechanism

The reason for the increase of $\delta$ might be the material bleaching.[15] If the molecules being in the excited state does not absorb (or weakly absorb) at the laser wavelength, the expression for the laser photon density S inside the solid ($h\nu \cdot S = dF_L / dx$) under the short pulse irradiation ($t_{L2} \ll \tau_{el}$) is

$$\frac{dS}{dx} = (1 - e^{-2\sigma S})\rho / 2 \qquad (6)$$

The absorption cross-section of hBN can be estimated from $\sigma = \alpha / C^* \approx 1.1 \cdot 10^{-15}$ cm², where $C^* = DOS \approx 1$ states/(eV·cell),[16] or $C^* = 1.8 \cdot 10^{20}$ cm$^{-3}$ in our experimental conditions. At low laser fluences the light penetration depth is the inverse of the "cold" material absorption coefficient: $\delta = 1/\alpha \approx 50$ nm. At higher fluences ($F_L > F_L^* = h\nu/\sigma$), the excitation distribution has to be taken into account. According to (6), S slowly decreases with x following a linear law and $\delta \approx 2F_L/(h\nu C^*)$. We see that under our experimental conditions the saturation fluence is $F_L^* = h\nu C^*/\alpha = 0.7$ mJ/cm². A formation of the bleaching zone near the sample surface than can be expected at laser fluences $F_L \geq F_L^*$. This fluence is in the range, where the UV2/UV1 intensity ratio increases.

We have calculated the dependence of the UV2-band intensity as a function of the laser fluence using the expression for the excited state density distribution in the solid at a distance $x$ from the surface.[15] After a passage of the short laser pulse the free exciton concentration is:

$$C_{exc}(x) = C^*/2 \cdot \left\{1 - \exp\left(2h\nu S(x)/F_L^*\right)\right\} \qquad (7)$$

where $x$ is a distance from the surface. Moreover, the probability of the bulk exciton localization can be expressed as:

$$P_{bulk}(x) = 1 - \exp(-x/l_{exc}), \qquad (8)$$

where $l_{exc}$ stands for a mean exciton displacement before localization (diffusion length). Correspondingly, a probability $1 - P_{bulk}(x)$ characterizes the surface exciton trapping. The relative intensity of the UV2 band (bulk emission) is then:

$$I_{UV2} = \int_0^\infty C_{exc}(x) P_{bulk}(x) dx \qquad (9)$$

that can be numerically solved by combining Eqs (6)-(9). The only adjusted parameter $l_{exc} = 70$ nm was obtained by comparison with the experiment. The theoretical curve shown by a solid line in Fig. 6 agrees well with the experiment. The exciton diffusion length of 70 nm is much longer than the range of the Förster-Dexter transport $l_{FD} \sim 2$ nm (see in Ref. 17). It may be due to the high energy of the laser photons, which is above the band-gap energy of hBN: $h\nu_L > E_g = 4.02$ eV.[9] The exciton diffusion length is known to increase with increasing photon energy.[18] Indeed, we have observed that the UV2 band is more intense when exciting at 5 eV then at 4 eV.[9] On the other hand, $l_{exc}$ depends on the sample preparation, and $l_{exc} >> l_{FD}$ may characterize a good crystallinity of the hBN sample.

The above discussion is valid (strictly speaking) for the semi-infinite solid only, which is not the case. Our samples are compacted of micro-crystals with mean size of ~3.1 μm. However, the excitation penetrates only in a small part of the crystal, which can be considered in this respect as a semi-infinite one, at least for moderate fluence levels (<20 mJ/cm²).

Recently, Hirayama et al in Ref. 12 have reported the effect of the nanosecond and femtosecond laser illumination on the cBN above the ablation threshold using near-IR lasers. They concluded that in contrast to the nanosecond laser, femtosecond laser processing keeps the chemical nature of the material unchanged. Such a result is confirmed by our observations.

## IV. CONCLUSION

In summary, the effect of nanosecond and femtosecond UV laser irradiations of graphite-like hexagonal boron nitride has been studied. The surface darkening of the sample has been observed after the laser treatment, which has been attributed to an enrichment of the surface layer by elemental boron. The ESCA analysis of the laser- treated samples and *in situ* fluorescence analysis of the plasma plume at laser energy above the ablation threshold (78 mJ/cm² under *fs*-laser pulses) confirmed this conclusion. The phase h-c transformation has not been evidenced in the present studies.

The laser-excited fluorescence spectra of hBN were found strongly modified under the laser processing. We distinguished two fluorescence bands assigned to the bulk (300-350 nm) and surface (370 nm) centers and considered their intensity ratio. In the nanosecond mode the intensity of the band due to the bulk centers progressively increases whereas the surface darkens. The initial spectrum recovers, with typical time is about $6.7 \cdot 10^3$ sec at residual pressure of $10^{-7}$ mbar in the sample chamber. This effect has been tentatively attributed to the hBN dissociation following by the oxidation of elemental boron in the surface layer. Boron oxide is transparent in the visible and protects deeper boron layers against oxidation. This explains the residual coloration of the sample surface. We found that in the case of nanosecond laser irradiation the darkening kinetics depends on the laser doze (typical value is ~40 J/cm²), whereas the laser intensity varied by a more than the order of magnitude.

Surface processing of hBN reveals that much lower laser energy is required to observe similar effect using femtosecond pulses as compared to nanosecond ones. We show that the femtosecond kinetics is not a function of the laser doze (or of the processing time) but it mainly depends on the laser fluence of a single pulse. The spectral changes are completely reversible: the low energy fluorescence spectra were immediately recovered when the laser fluence is decreased. This fact indicates that in this case the net effect is not related to the surface bond breaking or to the surface cleaning from the oxidizing species. We relate the effect to a sample bleaching by laser photons. The mechanism of this bleaching is discussed and quantitative agreement between the model and the experiment is found using the exciton diffusion length $l_{exc}$ =70 nm.

**ACKNOWLEDGMENTS**

This work was supported by the ULF-FORTH 00260 Contract of the European Commission. The authors are particularly grateful to F. Paillou (L.M.P. Poitier) and R. Flamand (I.M.P. Odello) for the ESCA measurements.

**Tables**

Table I          Impurity content of hBN sample (wt. %) determined by mass-spectrometry,
                 spectral analysis and X-ray microanalysis.

| C | O | Si | Ca | Al | Mg | Fe |
|---|---|----|----|----|----|----|
| 0.04±0.01 | 0.06±0.02 | 0.03±0.006 | 0.02±0.005 | 0.015±0.003 | 0.0006±0.0001 | 0.0025±0.0001 |

**Figure captions**

Fig. 1   Fluorescence spectra of the hBN sample treated with different pulse numbers N. The laser fluence is ~70 mJ/cm² in nanosecond pulse duration.

Fig. 2   SEM images of the initial hBN sample (a) and sample after a long irradiation by the KrF laser (b).

Fig. 3   Fluorescence spectra of the plasma plume in the laser ablation experiment: initial (solid line) and irradiated (dashed line) samples.

Fig. 4   Doze dependence of the normalized intensity of the UV2 fluorescence bands .

Fig. 5   Fluorescence spectra of the hBN sample treated with femtosecond laser pulse of different fluences of 0.7 mJ/cm² (a), 4.3 mJ/cm² (b) and 16.2 mJ/cm² (c). The ablation threshold is ~78 mJ/cm².

Fig. 6   Fluence dependence of the relative intensity of the UV2 fluorescence band. Solid circles (•) correspond to spectra obtained under irradiation; triangles (▲) correspond to spectra measured immediately after processing with a very small fluence of 0.7 mJ/cm². The solid line shows the calculated dependence using the exciton diffusion length of 70 nm.

Fig.1

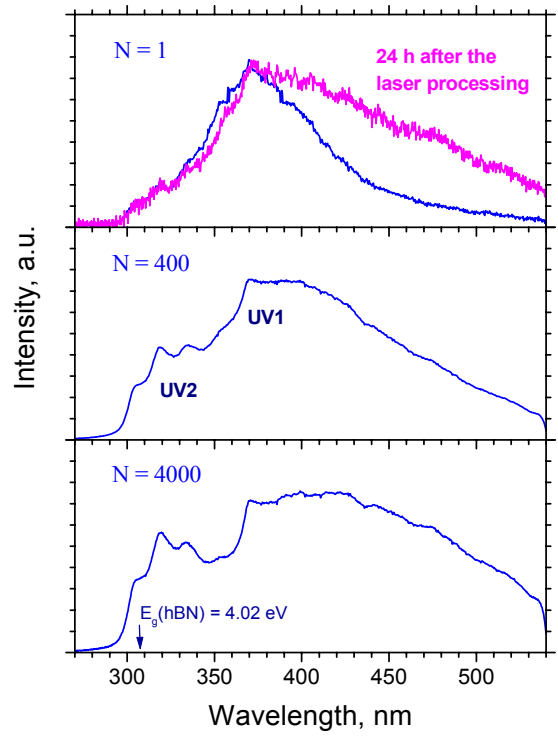

Fig. 2

a

10 µm

b

10 µm

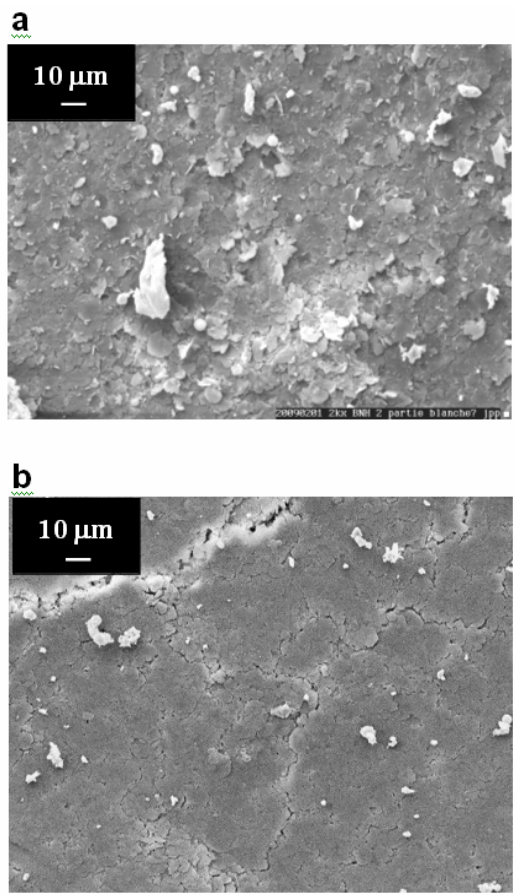

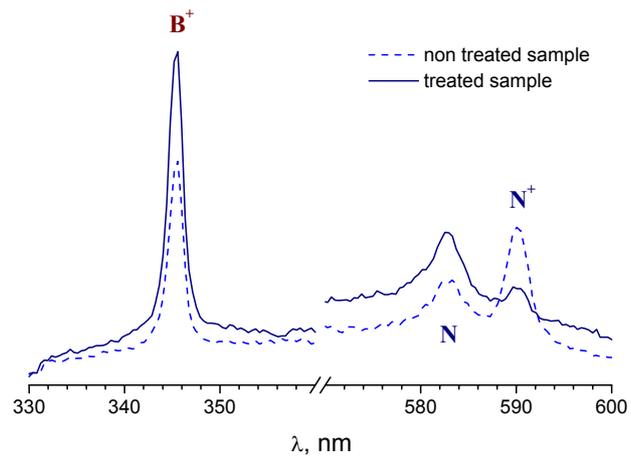

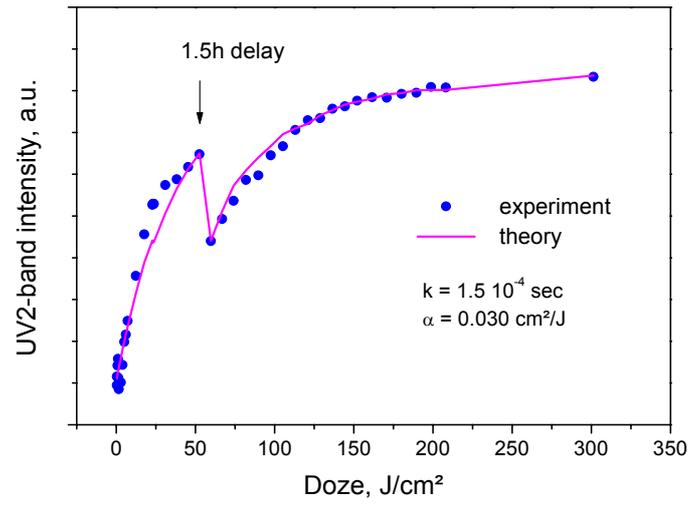

Fig.5

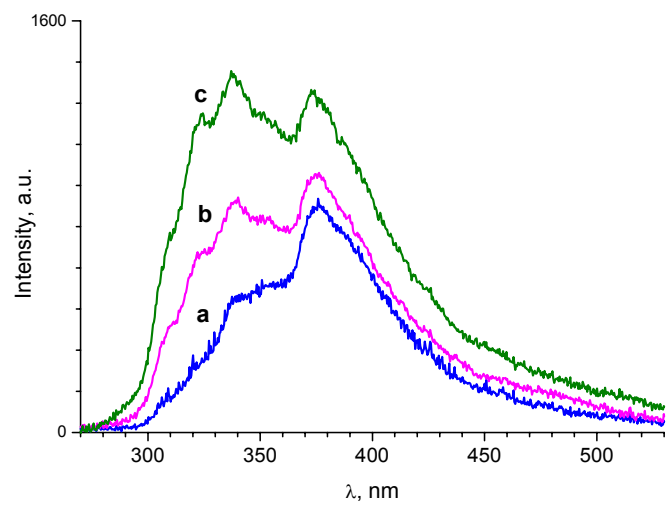

Fig.6

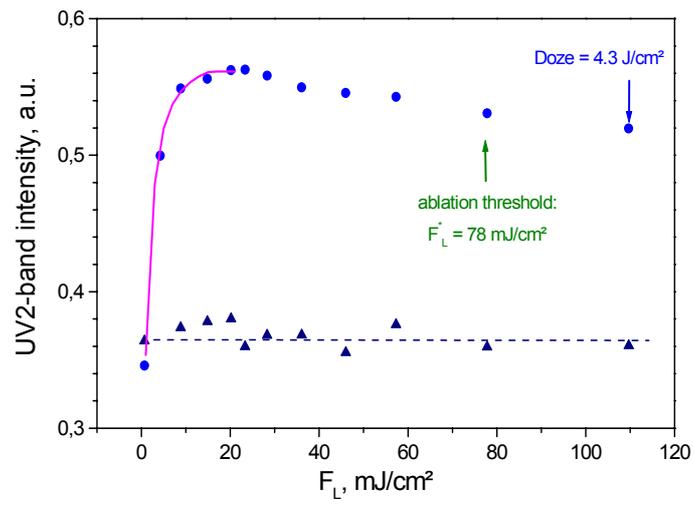